\documentclass[twocolumn,tightenlines,prd,floatfix,superscriptaddress,nofootinbib]{revtex4}

\usepackage{epsfig} 
\usepackage{amssymb} 
\usepackage{amsmath} 
\usepackage{amsfonts}

\newcommand{\be}{\begin{equation}}
\newcommand{\ee}{\end{equation}}

\newcommand{\vev}[1]{ \left\langle {#1} \right\rangle }

\newcommand{\D}[2]{\frac{d{#1}}{d{#2}}} 
  
\newcommand{\sfrac}[2]{#1/#2}

\begin{document}

\title{High energy neutrinos from charm in astrophysical sources}

\author{Rikard Enberg}
\thanks{Present address:\ Department of Physics and Astronomy, Uppsala University, Sweden}
\affiliation{Department of Physics, University of Arizona, Tucson, AZ 85721}

\author{Mary Hall Reno}
\affiliation{Department of Physics and Astronomy, University of Iowa, Iowa City, IA}

\author{Ina Sarcevic}
\affiliation{Department of Physics, University of Arizona, Tucson, AZ 85721}
\affiliation{Department of Astronomy, University of Arizona, Tucson, AZ 85721}

\begin{abstract}
Charm production gives rise to a flux of very high energy neutrinos from
astrophysical sources with jets driven by central engines, such as gamma ray
bursts or supernovae with jets. The neutrino flux from semi-leptonic decays of
charmed mesons is subject to much less hadronic and radiative cooling than the
conventional flux from pion and kaon decays and therefore has a dominant 
contribution at higher energies, of relevance to future ultrahigh energy 
neutrino experiments.
\end{abstract}

\maketitle

\section{Introduction}

Large underground or underwater experiments like IceCube~\cite{Ahrens:2003ix}
and KM3NeT~\cite{Katz:2006wv} are designed with the goal of observing 
high energy neutrinos produced in
astrophysical sources. The highest energy neutrinos, 
with energies of
 $10^{9}$~GeV and higher 
 may be observed in 
radio detection experiments 
\cite{radio}, 
and with an even higher energy threshold of 
$10^{12}$~GeV 
with  
acoustic detection experiments 
~\cite{Vandenbroucke:2004gv}. 
We consider astrophysical sources driven by a relativistic jet outflow,
accelerated by a central engine such as a black
hole~\cite{Woosley:1993wj,grbreview}. Shock accelerated protons in the jet outflow
may give rise to a high-energy neutrino flux~\cite{Waxman:1997ti}. These
neutrinos
are potentially produced in hadronic interactions: proton--proton 
 interactions 
 produce charged pions and kaons which subsequently
decay into muons and neutrinos. 
Above the threshold for $\Delta^+$ production, proton interactions with 
ambient photons also produce charged pions, and at higher energies,
kaons. 
The relative importance of the $pp$ and $p\gamma$ contributions to the
neutrino fluxes depends on the characteristics of the astrophysical environment. 
Several types of astrophysical sources have been studied in 
e.g.~\cite{Waxman:1997ti,RMW,Ando:2005xi,Koers:2007je,Meszaros:2001ms,Razzaque:2003uv,Horiuchi:2007xi,Wang:2008zm,Murase:2008sp,kt1,kt2}.

High energy pions and kaons are relatively long-lived and therefore subject to 
both hadronic and radiative cooling before they decay, which downgrades the neutrino
energies. Charm production and decay in astrophysical jets is also a source of
neutrinos \cite{kt1}.
In this paper, we show that production of charmed mesons in $pp$ 
collisions gives a large contribution to the neutrino flux at the highest
energies, since high energy charmed hadrons ($D^\pm, D^0$) have 
short lifetimes and therefore predominantly decay before they interact.  
Moreover, since the amount of
radiative cooling scales as $m^{-4}$, the larger masses of the
charmed hadrons lead to less cooling. 
The neutrino flux from charm is
therefore less suppressed up to higher energies. Even though the production
cross section is orders of magnitude smaller than for pions and kaons, 
neutrinos from charm decays become the 
dominant contribution at high energies.   The energy at which charm begins to 
dominate depends on the detailed properties of the astrophysical source, and the 
extent to which charm contributes depends on the maximum energy of the
accelerated protons in the jet.  

An example of an astrophysical source model is the 
slow-jet supernova (SJS) model, proposed  by 
Razzaque, M\'esz\'aros and Waxman  (RMW) \cite{RMW}. It is 
 characterized by a mildly relativistic jet propagating in a collapsing
star. This jet does not emerge from the source, and is sometimes referred to as a 
`choked jet.'    
This environment has a large optical depth \cite{Ando:2005xi,Koers:2007je}, so 
 neutrinos may be the only high energy signals to emerge.
On the other hand, the jet in a Gamma Ray Burst (GRB) 
(see e.g.~\cite{grbreview} for a review) is highly relativistic and the optical depth 
might be such that photons escape during the burst and do not thermalize.  
Other GRBs may be choked, so that only neutrinos escape~\cite{Meszaros:2001ms}.

We present here a new analysis of the neutrino flux for two types of source environments: 
the SJS and a GRB model, both of which have large magnetic fields in the jets.
Our treatment of the proton--proton collisions accounts for charmed meson
production, as well as pion and kaon production. The shorter lifetimes of charmed 
mesons  allow this channel to potentially dominate the neutrino
flux from $pp$ collisions.

\section{Astrophysical Sources}

Cooling times for protons and mesons depend on the proton and 
 photon densities in the jet, which are determined from the 
characteristics of the environment.  
 The internal shock radius of the jet is given by $r_j
= 2\Gamma_j^2 ct_v$, where $\Gamma_j$ is the bulk Lorentz gamma factor of the jet
outflow and $t_v$ is the variability time scale of the central engine. The
proton and electron number densities can be calculated in terms of the jet
luminosity, $L_j$, and 
the jet opening angle, 
$\theta_j$, as 
\be
n'_e = n'_p= \frac{L_j}{2\pi \theta_j^2 r_j^2 \Gamma_j^2 m_p c^3}.
\ee  
The primes denote quantities in the frame comoving with
the jet. The average magnetic field generated in the jet is given by 
\be
B' = \left(\frac{4\epsilon_B L_j}{\theta_j ^2 r_j^2 \Gamma_j^2 c}\right)^{1/2},
\ee 
where $\epsilon_B$
is the fraction of kinetic jet energy converted into magnetic fields. 
The comoving energy density of photons in the jet is 
\be
U_\gamma' = \frac{\epsilon_e L_j}{2\pi \theta_j^2 r_j^2 \Gamma_j^2 c},
\ee
 where $\epsilon_e$ is the fraction
of the energy radiated into photons. 

Depending on the optical depth, in some astrophysical sources photons may 
be thermalized.  
For a thermal photon distribution with temperature $T'$, the photon
number density is given by $n_\gamma' \propto (kT')^3$, leading to 
\be
n_\gamma' = \frac{2\zeta(3)}{\pi^{7/2}}
\left(\frac{15 U_\gamma'}{\hbar c}\right)^{3/4}.
\ee 
In the astrophysical environments where photons are not thermalized, the photon energies follow
a broken power-law spectrum~\cite{Band:1993eg}, characterized by the break energy
$E'_{\gamma,b}$, taken as a free parameter, and 
$U_\gamma'=n_\gamma' E_\gamma'$. 
 
\subsection{Slow Jet Supernova Model}

The RMW model of this system has a bulk jet
Lorentz factor $\Gamma_j=3$. The characteristic time variability of
the source is $t_v=0.1$~s, with a jet luminosity of $L_j=3\times
10^{50}$~erg/s and jet angle of $\theta_j\sim 1/3$. 
We take the energy fractions $\epsilon_{e,B}=0.1$.  
With these parameters, 
 the estimated size of the jet is on the order of  $10^{11}$ cm,
below a typical stellar radius for a pre-supernova star. 
Due to the large optical depth, photons thermalize.  

\subsection{Gamma Ray Bursts}

We also consider a generic case of a GRB. There is a wide range of
possible parameters and types of progenitors of GRBs, and the choice of
parameters we make here is to be considered as just one example of a
GRB-like jet.  
 The jet gamma factor can vary over a large range of
values, and the medium can be optically thick or thin. We choose the case of a
 GRB where the photons are not in thermal equilibrium and we take the
values $\Gamma_j = 100$, $L_j = 3\times 10^{50}$~erg/s,
$\theta_j=0.1$,
$t_v=10^{-3}$~s. We fix the photon break energy to be
$E'_{\gamma,b}=2.5$~keV~\cite{Preece:1999fv} and the spectral indices to be $-1$
below and $-2.25$ above the break energy. We take the same energy fractions, 
$\epsilon_{e,B}$, 
as for the SJS model.  

In Table~\ref{table:param2} we show the number densities of protons and photons,
the values of the magnetic field and the characteristic photon energy,
in the comoving frame, for the SJS and GRB models.

\begin{table}
\caption{The jet bulk Lorentz factor $\Gamma_j$, and
the comoving number densities of protons $n'_p$ and photons 
$n'_\gamma$, average photon energy $E_\gamma'$, 
and magnetic field in the jet $B'$ for the 
slow-jet supernova (SJS) and
gamma ray burst (GRB) models.}
\centering
\begin{tabular}{lcccccc}
\hline\hline
Source & $\Gamma_j$ & $n'_p$ [cm$^{-3}$] & $B'$ [G] & $E_\gamma'$ [keV]  
& $n_\gamma '$ [cm$^{-3}$] \\
SJS &3& $3.6\times 10^{20}$ & $ 1.2\times 10^{9}$  & $4.5$
& $2.8 \times 10^{24}$   \\
GRB &100& $3\times 10^{16}$ & $ 1.1\times 10^7$& $2.5$ & $1.1\times 10^{21}$
 \\[1ex]
\hline
\end{tabular}
\label{table:param2}
\end{table}

\section{Time and distance scales}

The neutrino flux is determined by the flux of accelerated protons,
the production of mesons, meson interactions and meson decay. This all occurs in
an environment of magnetic fields, photons, and a non-relativistic
baryon density. To begin with, we consider the characteristic time scales and
distance scales relevant to the accelerated protons. Since many of the processes
are relevant to mesons as well, we also indicate the application to mesons.

Protons are accelerated by Fermi shock acceleration. The acceleration time is
\begin{equation}
t'_{\rm acc} \simeq 3\times 10^{-12}\ {\rm s} \; \frac{\kappa}{10}\frac{0.3}{\epsilon_B^{1/2}}
\left(\frac{E_p'}{\rm GeV}
\right) \left(\frac{B'}{10^9 \,{\rm G}}\right)^{-1} \ ,
\label{eq:tacc}
\end{equation}
where the the parameter $\kappa$ is inversely related to the diffusion
coefficient. The details of the shock acceleration, in particular the orientation of the magnetic field relative to the shock, determine the value of diffusion coefficient \cite{Jokipii}.
Using the standard choice of $\epsilon_B=0.1$ and $\kappa =10$, the
acceleration time
converted to a distance is
\begin{equation}
L_N^\text{acc}= 10^{-2} \ {\rm cm}\ \left(\frac{E_p}{{\rm GeV}}\right)\left(\frac{B'}{10^9 \,{\rm G}}\right)^{-1}\ .
\end{equation}
A value of $\kappa =10$ taken here could be reduced by a factor of 10
or more depending on the orientation of the magnetic field \cite{Jokipii,prothstanev}. We comment below on the implications of
a smaller value of $\kappa$. 

Until proton cooling times (or proton interaction lengths) are smaller
than the acceleration time, the proton energy spectrum within the jet is
characterized by a power law which we take
to be $\phi_N\sim (E_p^{\prime})^{-2}$. We neglect
shock acceleration of the charged mesons to simplify the evaluation \cite{Koers:2007je}.

The time scales for proton cooling depend on the comoving
energies, particle densities, cross sections, and
magnetic fields. Generically, the cooling time is 
$t_\text{cool}' = E'/|dE'/dt'|$.
One can approximate $|dE'/dt'| \simeq c n'\sigma \Delta E'$,
where $\Delta E'$ is the particle energy loss 
and $\sigma$ is the cross section of the
hadron scattering with particles with comoving number density $n'$. For
scattering on photons, we use $\vev{ n'\sigma v}$, averaged over the photon spectrum.

There are several processes that contribute to the proton and/or meson cooling.  
 The effect of the energy loss,
accounting for the weighting by the incident flux,
can be described by a
cascade equation written in terms of $Z$-moments \cite{gaisser}. 
For the proton flux ($\phi_N$) the propagation
over distance $x'$ in the co-moving jet frame is given by 
\be
\left. \frac{d\phi_N}{dx'}\right|_{\rm cool} \!\!\!= 
-\frac{\phi_N}{\ell_N^{\rm had}}
+ Z_{NN}^{\rm had} \frac{\phi_N}{\ell_N^{\rm had}}
+\dots 
\ee
where $\ell_N ^{\rm had} = (\sigma_{pp}n_p')^{-1}$ describes interactions 
with target protons. The ellipsis
signifies other cooling mechanisms, such as  $p\gamma$ reactions, and electromagnetic cooling, to be discussed below.
The $Z$-moment is defined by
\begin{equation}
Z_{NM} = \int_0^1 dx_E x_E^{\alpha-1} \frac{dn_{N\rightarrow M}}{dx_E}
\label{Zintegral}
\end{equation}
where the flux of $N$ is $\phi_N\sim (E')^{-\alpha}$,
$x_E\equiv E'_M/E'_N$ and $dn/dx_E$ describes the cross section normalized
energy distribution of the meson $M$ produced by $N$ (or from $N$ decay).

Using the comoving 
proton number density from 
Table~\ref{table:param2} and the
cross section $\sigma_{pp} \simeq 5\times 10^{-26}\ {\rm cm}^2$, 
the interaction
length is $\ell_N^{\rm had}\simeq 5.5\times 10^4$ cm for the SJS model of RMW,
and $\ell_N^{\rm had}\simeq 6.8\times 10^{8}$ cm for the GRB.  
We used these same interaction lengths
for the hadronic interaction of the mesons.  Using the exact value for the 
inelastic cross section for mesons does not affect the result because 
the hadronic cooling time for mesons is much larger than its radiative cooling times. 

The distance scale corresponding to the hadronic
cooling time is 
\be
L_N^{\rm had} = \frac{\ell_N^{\rm had}}{1-Z_{NN}^{\rm had}}\simeq ct_N^{\rm had}.
\ee  
The $Z$-moment takes the inelasticity into account, and is equivalent to
setting
$\sfrac{\Delta E'}{E'} \simeq 1-Z_{NN}^{\rm had}$
for this hadronic scattering process.  
We will use $Z_{NN}^{\rm had}=0.5$~\cite{tig}.

In addition to hadronic interactions, protons can lose energy via 
photon interactions such as inverse Compton (IC) scattering, where the proton
loses energy to the ambient photons, or through inelastic $p\gamma$ interactions
if the proton energy is above the $\Delta^+$ threshold.
For the IC process, as long as the center of mass energy 
 is small compared to the proton mass, one can use the Thomson cross section,
$\sigma_{Tp} = \sigma_T(\sfrac{m_e^2}{m_p^2})$.
This approximation is good for protons scattering off 
keV photons for energies up to $E'_p\sim 10^5$ GeV. Above this energy, 
IC becomes irrelevant.
The effective scattering length for inverse Compton scattering is 
\be
L_N^{IC} = \frac{3m_p^4c^4}{4 \sigma_{T} m_e^2 E_p' U_\gamma'}.
\ee
This effective scattering length is rescaled by $(m_M/m_p)^4$ for mesons.

\begin{figure}
\begin{center}
\epsfig{file=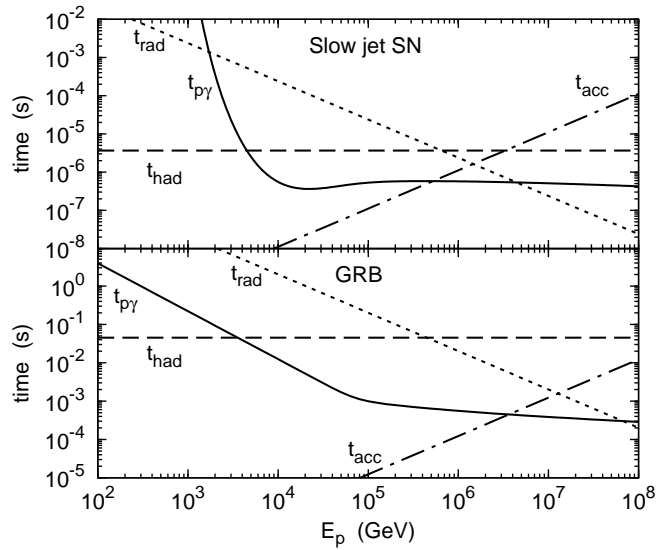,clip=,width=1\columnwidth}
\end{center}
\caption{Proton cooling times for SJS and for GRB in the frame co-moving with the jet.}
\label{fig:protontimes}
\end{figure}

For $E_\gamma' = 5$ keV, 
the threshold energy for $\Delta^+$ production in $p\gamma$ interactions is 
$E_{p,\text{th}}'=2\times 10^5$ GeV.  For $p\gamma$ scattering, the averaged reaction 
rate is given by, 
 
\be
\vev{n'\sigma v} = \frac{c}{8\beta'_p E_p'^2} 
\int dE_\gamma' \frac{\hat n_\gamma(E_\gamma')}{E_\gamma'^2}
\int ds  (s-m_p^2) \sigma_{p\gamma}(s),
\ee
where $\beta_p'$ is the proton beta factor, $s$ is the center-of-mass energy 
squared, and $\hat n_\gamma(E_\gamma') dE_\gamma'$ is the photon number density
in the energy range between $E_\gamma'$ and $E_\gamma'+ dE_\gamma'$. 
The cross section $\sigma_{p\gamma}$ is given by the resonance plus
continuum multiparticle production contributions given in Ref.~\cite{Mucke:1999yb}. 
The photon distribution $\hat n_\gamma$ is thermal for the SJS
and a power law for the GRB~\cite{Waxman:1997ti}.  
In determining the cooling time, 
$t_{p\gamma}'$, we take 
$\Delta E'/E'\approx 0.2$ \footnote{The Bethe--Heitler process $p\gamma\to p e^+
e^-$ is less effective than inelastic $p\gamma$  above the 
threshold~\cite{RMW} and we will
therefore neglect it.}.

Finally, there is a cooling distance associated with
synchrotron radiation because of the magnetic field
in the jet. This distance scale is
\be
L_N^{B} = \frac{6\pi m_p^4 c^4}{\sigma_{T}m_e^2E_p' B'^2}. 
\ee
Again, meson interaction lengths can be obtained by scaling by $(m_M/m_p)^4$.

The cooling times and the acceleration time for protons
are shown in Fig.~\ref{fig:protontimes}.
For the proton flux, until cooling becomes important, acceleration
dominates. 
For both the SJS and GRB example models, cooling through proton-photon
interactions eventually dominates the acceleration. The crossover between
acceleration time and $t_{p\gamma}$ in Fig. 1 occurs in an energy
regime where $t_{p\gamma}$ is nearly constant, although there is
strong energy dependence at lower energies. 

When cooling dominates acceleration, the proton flux is cut off.
Details of the cutoff depend on the energy dependence and mean 
inelasticity of the interaction \cite{prothstanev,protheroe}. Generically,
we can write
\begin{equation}
\phi_N(E) \equiv \phi_N^0(E) f_N(E)= AE^{-2}f_N(E) \ .
\end{equation}
We apply a smooth cutoff $f_N(E)$ of the form of Eq.~(20) in Ref.~\cite{prothstanev}
with a spectral index $\Gamma = 2$ and the parameter $\delta = 1$:
$$
f_N(E) = \left( 1+\left(\frac{E}{E_{\rm max}}\right)^\delta\right)
\exp\left[ -\frac{(\Gamma-1)}{\delta} \left(\frac{E}{E_{\rm max}}\right)^\delta\right]\ .
$$
The energy at which the cross-over of acceleration and cooling times occurs, in the
co-moving frame, is $E_\text{max}\simeq 5.2\times 10^5$
GeV for the SJS and $E_\text{max}\simeq 3.6\times 10^6$ GeV for the GRB model.   Our result for the SJS 
model is in agreement with previous estimates of proton cooling times \cite{RMW,Ando:2005xi}.

\begin{figure}
\begin{center}
\epsfig{file=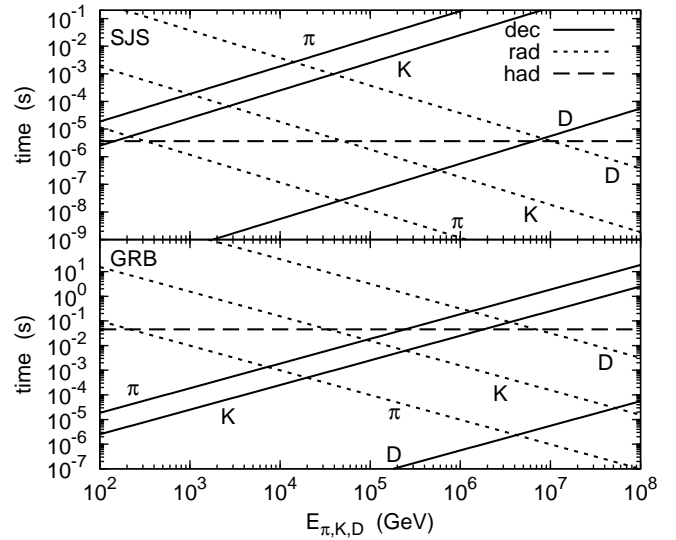,width=1\columnwidth}
\end{center}
\caption{Meson cooling times for SJS and GRB in the frame co-moving with
the jet.}
\label{fig:mesontimes}
\end{figure}

Fig.~\ref{fig:mesontimes} shows the meson cooling times and the decay times in the frame
co-moving with the jet for both the SJS and GRB models. 
For mesons, the decay length  and decay time includes the gamma factor (in the comoving frame), e.g.,
$L_M^{\rm dec} = (\sfrac{E'_M}{m_M c^2}) c \tau_M$.
The proper decay lengths $c \tau_M$ are listed in Table~\ref{table:zmesons}.
Pion and kaon cooling times are in agreement with previous 
 estimates \cite{RMW,Ando:2005xi}.  
One can see from Fig. \ref{fig:mesontimes} that for $D$ mesons, the
cross over of the decay time with the cooling time (hadronic or radiative)
occurs at much higher energies than for pions and kaons.

\section{Meson and neutrino fluxes}

Neutrinos are produced through meson production followed by meson decay. It is
convenient to define the combined cooling length $\tilde{L}_M$ for
meson $M$ through
\begin{equation}
\tilde{L}_M^{-1} = (L^{\rm had}_N)^{-1}+
\frac{m_p^4}{m_M^4}[(L_N^{IC})^{-1} +  (L_N^{B})^{-1}]\ ,
\end{equation}
including the mass rescaling factor.
An effective length for a given meson which also accounts for the decay is
$(L_M^{\rm eff})^{-1}=(\tilde{L}_M)^{-1}+ (L_M^{\rm dec})^{-1}$.
With this definition, the equation describing the propagation of mesons $M$
in the jet is
\begin{equation}
\frac{d\phi_M}{dx'} = -\frac{\phi_M}{L_M^{\rm eff}}+{Z_{NM}}
\frac{\phi_N}{\ell_N^\text{had}}+ Z_{NM}^\gamma \frac{\phi_N}{\ell_N^\gamma}\ ,
\label{meson}
\end{equation}
where $Z_{NM}$ are the $Z$-moments for $pp$ meson production and
$Z_{NM}^\gamma$ are the $Z$-moments for $p\gamma$ meson production. We have here assumed a
constant density that does not vary with $x'$. Finally, the neutrino flux from
meson $M$ is
obtained from the meson flux by integrating the equation
\be
\D{\phi_\nu}{x'} = Z_{M\nu} 
\frac{\phi_M}{L_M^{\rm dec}}\ ,
\ee
where $Z_{M\nu}$ is the decay $Z$-moment listed in Table~\ref{table:zmesons}. 
In the limit $x'\to\infty$ we have 
\begin{eqnarray}
\phi_\nu(E') =  Z_{M\nu} 
\frac{L_M^{\rm eff}}{L_M^{\rm dec}} 
 \frac{{Z_{NM}}\ell_N^\gamma + Z_{NM}^\gamma \ell_N^\text{had}}{\ell_N^\text{had}+\ell_N^\gamma}
\phi_N(E') \ .
\label{leptonflux} 
\end{eqnarray}
This has the correct limiting behavior, e.g., when decays dominate the meson effective length
$L_M^{\rm eff}$ and the energy is below the $p\gamma$ cross section threshold. In this limit,
\be
\phi_\nu(E')= Z_{M\nu}Z_{NM}\phi_N(E')\ ,
\ee
essentially the same as in Ref. \cite{Ando:2005xi} at low energy,
however, this form has a smoother transition to the high energy behavior.

We calculate the $Z$-moments for charm production 
 taking into account the
effects of parton saturation at high energies as in Ref.\ \cite{ERS}.  
These moments are energy dependent, with weak dependence at higher energies.   
In our calculation of the neutrino flux, for 
simplicity we use the constant value of the 
$Z$-moments for charm given in Table~\ref{table:zmesons}. We neglect the 
production of other charmed particles. We estimate that the inclusion of $\Lambda_c$ 
production would increase the flux of neutrinos from charm by $\sim 15\%$ \cite{prs}. 
For $\pi^\pm$ and $K^\pm$
production, however, we assume Feynman scaling, leading to energy independent 
$Z$-moments. 
 To compute
the $Z$-moments for $pp$ we use the parametrization of the 
rapidity distribution~\cite{Costa:1995gd}
\be
\frac{dN_\pi}{dx_E} = 0.12 \frac{(1-x_E)^{2.6}}{x_E^{2}}
\ee
where $x_E=E_\pi/E_p$, and for $p\gamma$ we fit this form 
to the data for $\pi^0$ production in Ref.\ \cite{Adloff:2000ua}. 
For $K^\pm$ we take the $\pi^\pm$ result rescaled by 
0.1~\cite{Alner:1987wb}.  
The decay $Z$-moments are calculated using the expressions in Refs.\
\cite{Lipari:1993hd,Bugaev:1998bi}.
The values for $Z$-moments are shown in
Table~\ref{table:zmesons}. 

\begin{figure}[tb]
\begin{center}
\epsfig{file=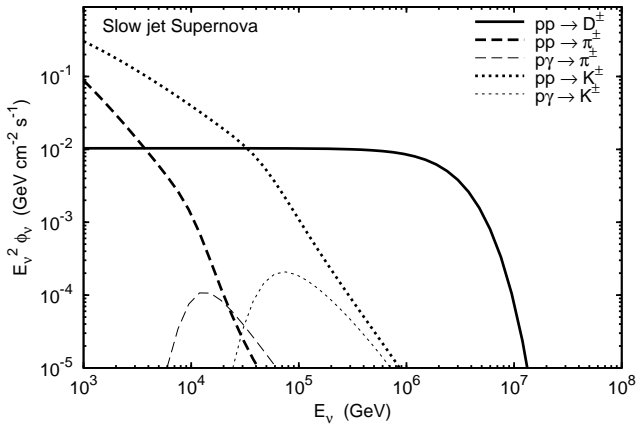,width=1\columnwidth}
\end{center}
\caption{Muon neutrino plus antineutrino flux from a slow jet supernova
 in the observer's frame, without including neutrino oscillation effects as they propagate on their way to Earth.} 
\label{fig:fluxesSJS}
\end{figure}

\begin{figure}[tb]
\begin{center}
\epsfig{file=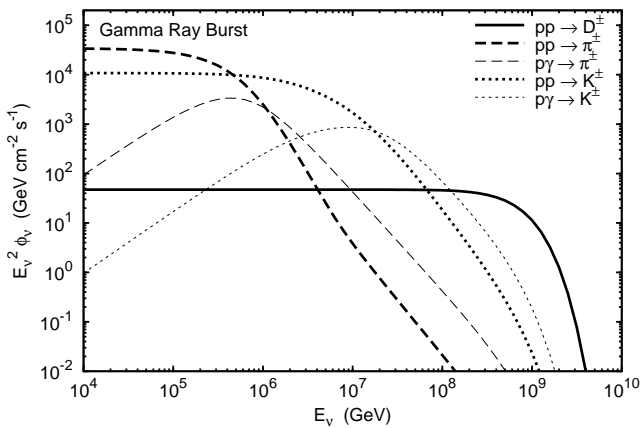,width=1\columnwidth}
\end{center}
\caption{Muon neutrino plus antineutrino flux from a gamma ray burst in the observer's frame, without including neutrino oscillation effects as they 
propagate on their way to Earth.} 
\label{fig:fluxesGRB}
\end{figure}

To evaluate the neutrino flux from a source in the Earth's frame, co-moving energies are
scaled by the jet bulk Lorentz factor $\Gamma_j$. The neutrino flux is related to the jet 
luminosity $L_j$, the distance to the source $d_L$, taken here to be $20 \text{ Mpc}=6.17\times 10^{25}$~cm,
and the jet opening angle shown in Table I. The neutrino flux is 
\begin{eqnarray}
\nonumber
\phi_\nu(E) &=& Z_{M\nu} 
\frac{L_M^{\rm eff}}{L_M^{\rm dec}(\ell_N^\text{had}+\ell_N^\gamma)} \\
\nonumber
&\times & ({Z_{NM}}\ell_N^\gamma + Z_{NM}^\gamma \ell_N^\text{had})\\
&\times&
\frac{L_j \Gamma_j^2}{2\pi\theta_j^2 d_L^2\ln (E'_{\text{max}}/E'_{\text{min}})} E^{-2}\ .
\end{eqnarray}
The quantity $E'_\text{max}$ is the energy where cooling and acceleration cross over, and
$E'_\text{min}=m_p c^2$ is the proton's rest energy.
In Fig. \ref{fig:fluxesSJS} we show the neutrino flux, $\phi_\nu$, as 
a function of neutrino energy for the slow jet supernova model, and
in Fig. \ref{fig:fluxesGRB}, for the gamma ray burst model. The fluxes shown are from $pp$ and $p\gamma$
interactions. At low energies, the dominant sources of neutrinos are
pions and kaons produced in $pp$ interactions.  Above
approximately $E\simeq 10^7$ GeV, $p\gamma$ 
production of pions and kaons becomes comparable in strength for
the GRB, while for the SJS it is subdominant for all energies.

\begin{table}[tb]
\caption{$Z$-moments and decay lengths used in the evaluation of the muon
neutrino plus antineutrino flux. The $p\rightarrow p$ moment is
taken to be $Z_{NN}^{\rm had}=0.5$. \label{table:zmesons}}
\begin{tabular}{lccccc}
\hline
$M$ & $Z_{M\nu}$ & $Z_{NM}$ & $Z_{NM}^\gamma$ &  $c\tau$ [cm] \\
\hline
$\pi^\pm$      & 0.061  &  0.55  & 0.13  & 780\\
$K^\pm$        & 0.19   &  0.055 & 0.0065 & 370\\
$D^\pm$        & 0.045  &  $2.4\times 10^{-3}$ & -- & $3.2\times 10^{-2}$\\
$D^0$          & 0.017  &  $5.6\times 10^{-3}$ & -- & $1.2\times 10^{-2}$\\
\hline
\end{tabular}
\end{table}

At low energies, where cooling is not important for the $D$ mesons,
the $D^+$ and $D^0$ contributions are approximately equal using the
parameters in Table~\ref{table:zmesons}.  At high energies, $pp$ production of charmed mesons, with their
shorter lifetimes, dominate although the maximum proton energy in the GRB jet
restricts the dominance of the neutrino flux from charm to a narrow energy interval.
As noted in Eq.~(\ref{eq:tacc}), the acceleration time depends on unknowns including $\kappa$. If $\kappa =1$,
the maximum proton energy in the comoving frame increases by a factor of 10, increasing the
range of energies for which charm decays contribute to the neutrino flux.

To discuss the potential for neutrino flux attenuation in the source,
we review the proton and neutrino optical depths for the slow jet
supernova model, where the jet does not emerge from the
source. The jet propagates to
$r_j\sim 5\times 10^{10}$ cm in the SJS model. The optical depth
$\tau'=\sigma n' r_j/\Gamma_b$
for protons in $pp$ and $p\gamma$ interactions are
\begin{eqnarray*}
\tau_{pp}'&\simeq & 3\times 10^5\\
\tau_{p\gamma}' &\simeq & 10^6\ {\rm above\ threshold}
\end{eqnarray*}
The proton-proton cross section is a factor of $10^8$ larger than the
neutrino cross section with protons at $E_\nu=10^6$ GeV. At
high energies, the cross section scales like $E_\nu^{0.44}$, so 
even at $E_\nu=10^8$ GeV, the proton cross section is a factor of
more than $10^7$ larger than the neutrino cross section.
At $10^6$ GeV, the optical depth for the neutrino within the jet
is 
$$\tau_{\nu p}' \simeq 2\times 10^{-2}\ .$$
Neutrino attenuation within the jet is therefore not important.

The densities inside and outside of the jet are not the same.
In the SJS model, the jet does not reach the stellar envelope, however,
$r_j$ is large enough that the relevant densities would be for
He or H. Taking as an example 
\cite{mr2001}
$\rho = (r_j/r)^3$ g/cm$^3$ for the He density,
the optical depth for a $10^6$ GeV neutrino is $\tau_{\nu p}\simeq 10$. 
On the other hand, if the relevant density is for the H envelope, the density
is much lower and the optical depth $\tau_{\nu p}\ll 1$. This would indicate that
some attenuation may be important in a full evaluation of neutrino fluxes
for these types of sources with a more detailed model of the stellar exterior \cite{Horiuchi:2007xi}.
For GRBs, the hadronic jet emerges, so given that the 
neutrino-nucleon cross section is so much lower than the proton-proton cross section, neutrino
attenuation in GRBs should not be important.

Taking into account the uncertainty in the acceleration differential 
index does not have a significant effect on  the relative importance
of the various contributions to the neutrino flux.  
The high energy  neutrino flux in the two examples shown here is dominated
by charmed meson decay. 
In contrast to pions and kaons, charmed mesons give equal amounts
of $\nu_e+\bar{\nu}_e$ and $\nu_\mu+\bar{\nu}_\mu$ at the source since the 
neutrinos are coming from semileptonic decays. 
This could give different flavor
ratios of neutrinos in neutrino telescopes at very high energies.  This has also
been noted in Ref. \cite{kt1}.
Since 
many properties of astrophysical sources are not well understood, 
neutrino measurements may hold the key to understanding the environments in
which they were produced.  
Charm production in astrophysical jets provides an important enhancement of the 
neutrino flux at very high energies.  

Current neutrino experiments, such as IceCube which has only partial implementation of the 
strings, have already put some limits on neutrino flux from a specific astrophysical source 
\cite{icecube_source}.  Once all the strings are implemented in IceCube, the 
 sensitivity for detection of the astrophysical neutrinos will improve significantly.  
In addition, future cubic kilometer deep-sea neutrino detector \cite{km3} will also have a very 
good chance to look for astrophysical neutrinos from point sources.

\begin{acknowledgments}
This research was supported by 
US Department of Energy 
contracts DE-FG02-91ER40664, DE-FG02-04ER41319 and DE-FG02-04ER41298.
\end{acknowledgments}


\end{document}